\documentclass[aps,prl,twocolumn,superscriptaddress]{revtex4-2}
\usepackage{amsmath,amssymb,graphicx}

\begin{document}

\title{Wave--Breaking Phenomena in Quark--Gluon Plasma}

\author{Biswarup Paul}
\affiliation{Lalit Narayan Mithila University, Darbhanga, 846004, Bihar, India}

\date{\today}

\begin{abstract}
We investigate the onset of wave--breaking in the quark--gluon plasma (QGP) formed in heavy--ion collisions at both RHIC and the LHC. A nonlinear longitudinal color field is coupled to a three--dimensional viscous hydrodynamic background constrained by experimental conditions: Pb--Pb at $\sqrt{s_{\rm NN}}=5.02$ TeV (LHC) and Au--Au at $\sqrt{s_{\rm NN}}=200$ GeV (RHIC). The instantaneous wave--breaking threshold is determined from the in-medium plasma frequency, while the field evolution follows a nonlinear Landau equation with Debye screening and expansion damping. At the LHC, the higher initial temperature and density drive the system above the threshold within $\tau \lesssim $ 2.3 fm/$c$, whereas at RHIC the cooler and more dilute medium delays wave--breaking to $\tau \simeq 2.8$~fm/$c$. Scans over initial temperature, density, thermalization time, QGP lifetime, and collision system (RHIC vs LHC) confirm the robustness of this instability against parameter variations. These results identify wave--breaking as a universal microscopic mechanism for the early loss of coherence and rapid onset of hydrodynamics in the QGP, providing a common explanation for the fast equilibration observed at both RHIC and the LHC.

\end{abstract}

\maketitle

\section{Introduction}

Understanding how the quark--gluon plasma (QGP) loses its initial coherence and approaches near--ideal hydrodynamic behavior within a few fm/$c$ is a central challenge in high--energy nuclear physics. Experimental data from LHC Pb--Pb collisions at $\sqrt{s_{\rm NN}}=5.02$ TeV (ALICE) and from RHIC Au--Au collisions at $\sqrt{s_{\rm NN}}=200$ GeV both indicate that collective flow builds up on timescales far shorter than kinetic theory alone would predict. This suggests the existence of rapid microscopic mechanisms that dissipate coherent field energy into bulk degrees of freedom. In electromagnetic plasmas, large--amplitude oscillations undergo wave--breaking once particle trajectories cross, leading to efficient energy transfer and fast isotropization~\cite{Dawson:1959,Maity2013,Romatschke2003,Gelis2010}. By extending this concept to QCD matter under realistic LHC and RHIC conditions, we provide a unified framework for understanding how chromodynamic fields in the early QGP can destabilize and seed the fast equilibration inferred from experimental observables. 
In particular, the predicted timescales for wave--breaking coincide with the experimentally observed onset of elliptic and higher--order anisotropic flow measured by ALICE at the LHC~\cite{ALICE:2016ccg} and by PHOBOS/PHENIX at RHIC~\cite{PHOBOS:2007dmu}, as well as with quarkonium suppression trends~\cite{ALICE:2024} and enhanced early thermal photon and dilepton emission~\cite{ALICE:2016fzo,PHENIX:2021nib}, offering multiple testable signatures of this mechanism in heavy--ion collisions.

\section{Theoretical Framework}

The space-time evolution of the QGP created in ultrarelativistic heavy-ion collisions can be modeled by a boost-invariant, longitudinally expanding background with approximate local thermalization \cite{Bjorken:1982qr,Heinz:2013th}. Under isentropic expansion, the temperature decreases as 
\begin{equation}
T(\tau) = T_0 \left(\frac{\tau_0}{\tau}\right)^c, 
\label{eq:temp}
\end{equation}
where $\tau_0$ is the thermalization time, $T_0$ the initial temperature, and $c$ depends on the expansion profile ($c=1/3$ for ideal Bjorken scaling). The parton density evolves as 
\begin{equation}
n(\tau) = n_0 \left(\frac{\tau_0}{\tau}\right)\left(1+\frac{\tau}{t_r}\right)^{-2},
\label{eq:density}
\end{equation}
where $t_r$ is the relaxation timescale accounting for collisional broadening and medium response \cite{Baier:2000mf, Kurkela:2018qeb}.

The maximum sustainable chromoelectric field before wave--breaking is given by the balance between collective restoring forces and quark acceleration, yielding \cite{Dawson:1959, Akhiezer:1975}
\begin{equation}
E_{\text{WB}}(\tau) \sim \frac{m_{\text{eff}}\omega_p(\tau)}{g_s C_R},
\label{eq:ewb}
\end{equation}
where $m_{\text{eff}}$ is the in-medium quasiparticle mass and $C_R$ the quadratic Casimir of the representation ($C_R$ = 4/3 for quarks). The plasma frequency in a relativistic QGP is
\begin{equation}
\omega_p^2(\tau) \sim \frac{g_s^2 n(\tau)}{3T(\tau)} ,
\label{eq:plasma}
\end{equation}
analogous to the QED case but modified by QCD color factors and the effective thermal distribution of quarks and gluons \cite{Blaizot:2001nr, Lebedev:1989ev}. 

The growth and nonlinear saturation of collective chromoelectric fields can be captured by a Ginzburg–Landau–type evolution equation, 
\begin{equation}
\frac{dE}{d\tau} = \left[\kappa m_D(\tau) - \alpha_0 \frac{\tau_0}{\tau}\right]E - \beta E^3,
\label{eq:evolution}
\end{equation}

where $m_D(\tau)$ = $g_s T(\tau)\sqrt{1+N_f/6}$ is the Debye screening mass, $\kappa$ quantifies plasma instabilities, $\alpha_0$ encodes longitudinal dilution, and $\beta$ represents nonlinear self-interaction of the field. The values of $\kappa$, $\alpha_0$ and $\beta$ are 0.55, 0.7 and 1, respectively \cite{Mrowczynski:1993qm, Romatschke:2003ms}. 

Equations (\ref{eq:temp})–(\ref{eq:evolution}) provide a consistent framework to analyze the nonlinear amplification and eventual breaking of color fields in the evolving QGP, connecting hydrodynamic expansion, plasma instabilities, and wave--breaking thresholds. These relations form the basis for the quantitative results presented below.


The initial conditions are constrained by global hydrodynamic analyses and bulk hadron spectra at the LHC (Pb--Pb, $\sqrt{s_{\rm NN}}=5.02$ TeV) and RHIC (Au--Au, $\sqrt{s_{\rm NN}}=200$ GeV)~\cite{ALICE:2016fzo,ALICE:2022wpn,Gale:2013da,Heinz:2013th,PHENIX:2004vcz,STAR:2008med}.  Representative ranges are summarized in Table~\ref{tab:params}.  For quantitative estimates we employ central values, corresponding to $T_{0}\!\simeq\!0.5$ GeV, $n_{0}\!\simeq\!6.5$ fm$^{-3}$, and $\tau_{0}\!\simeq\!0.4$ fm/$c$ at the LHC, and $T_{0}\!\simeq\!0.32$ GeV, $n_{0}\!\simeq\!3$ fm$^{-3}$, and $\tau_{0}\!\simeq\!0.8$ fm/$c$ at RHIC, with lifetimes $\tau_{f}\!\sim\!10$ and $7$ fm/$c$, respectively.  

\begin{table}[h]
\centering
\caption{Representative ranges of initial conditions extracted from hydrodynamic modeling and hadronic spectra.}
\label{tab:params}
\begin{tabular}{lcc}
\hline\hline
Parameter & LHC & RHIC \\
\hline
$T_0$ [GeV] & $0.45$--$0.55$ & $0.30$--$0.35$ \\
$n_0$ [fm$^{-3}$] & $5$--$8$ & $2.5$--$4$ \\
$\tau_0$ [fm/$c$] & $0.2$--$0.6$ & $0.6$--$1.0$ \\
$\tau_f$ [fm/$c$] & $8$--$12$ & $6$--$8$ \\
\hline\hline
\end{tabular}
\end{table}

\begin{figure}[t]
\includegraphics[width=\linewidth]{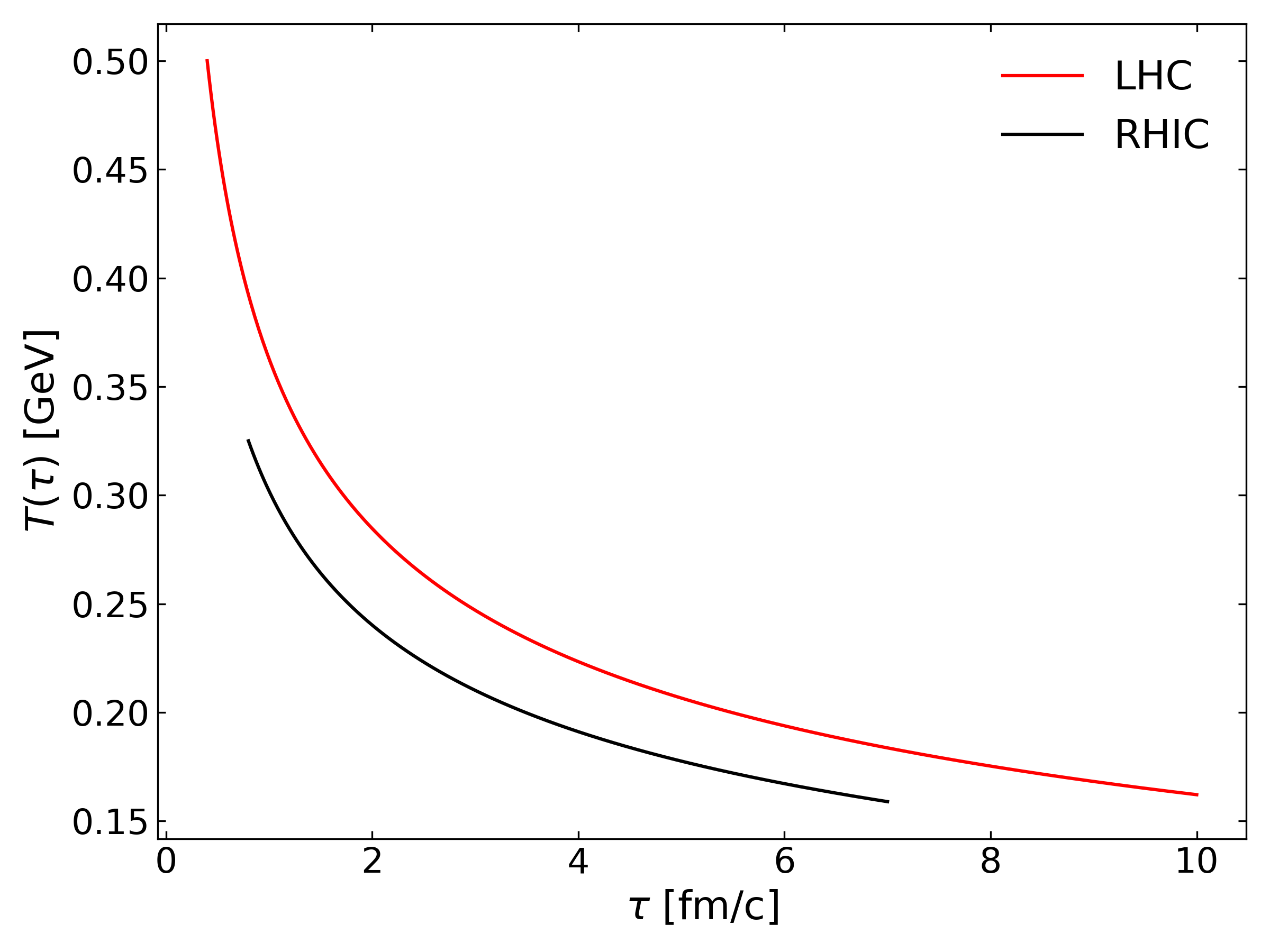}\vspace{3pt}
\includegraphics[width=\linewidth]{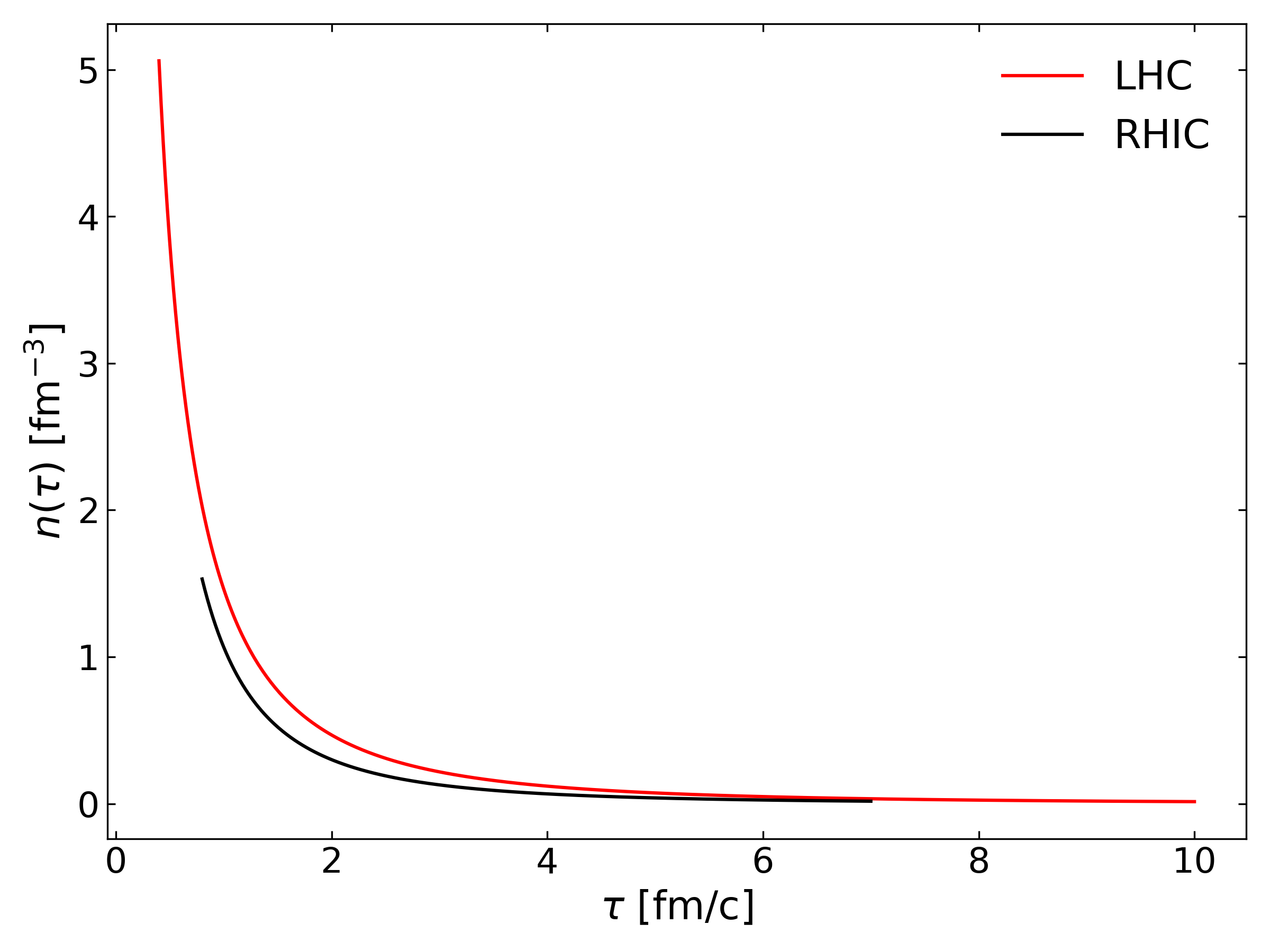}
\caption{
Time evolution of temperature $T(\tau)$ (top) and number density $n(\tau)$ (bottom) in Pb--Pb collisions at $\sqrt{s_{\rm NN}}=5.02$ TeV at LHC (red) and in Au--Au collisions at $\sqrt{s_{\rm NN}}=200$ GeV at RHIC (black).}
\label{fig:Ttau}
\end{figure}


\begin{figure}[t]
\includegraphics[width=\linewidth]{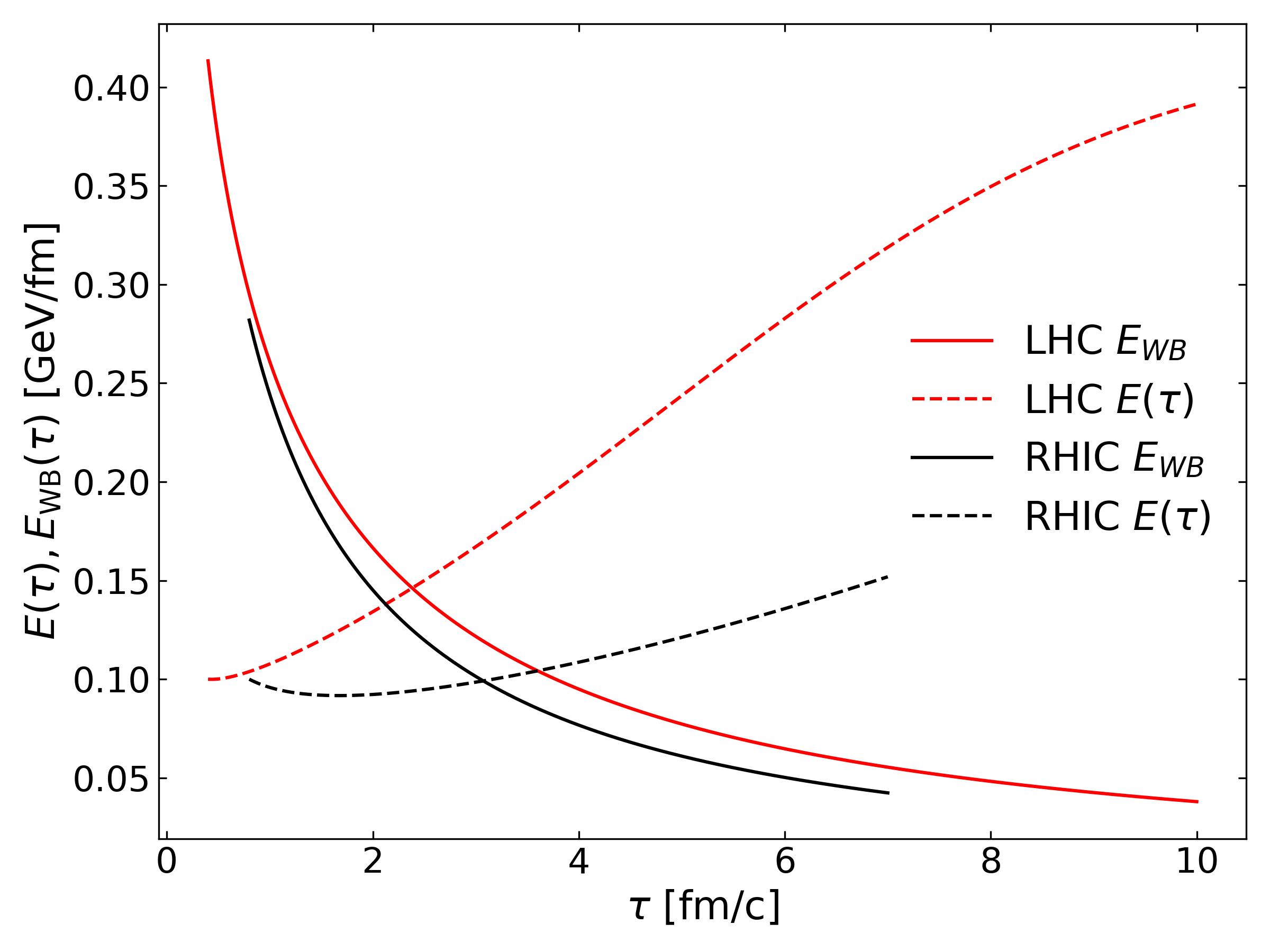}
\caption{Time evolution of the chromoelectric field $E(\tau)$ (dashed) compared with the wave--breaking limit $E_{\rm WB}(\tau)$ (solid) for Pb--Pb collisions at $\sqrt{s_{\rm NN}}=5.02$ TeV at LHC (red) and in Au--Au collisions at $\sqrt{s_{\rm NN}}=200$ GeV at RHIC (black). The crossing of $E(\tau)$ and $E_{\rm WB}(\tau)$ indicate the onset of wave--breaking in the QGP.}
\label{fig:Ewb}
\end{figure}

\section{Results and discussions}

Figure~\ref{fig:Ttau} shows the time evolution of the temperature $T(\tau)$ and number density $n(\tau)$ for Pb--Pb collisions at the LHC ($\sqrt{s_{\rm NN}}=5.02$ TeV) and Au--Au collisions at RHIC ($\sqrt{s_{\rm NN}}=200$ GeV). The initial conditions (Table~I) reflect values extracted from hydrodynamic modeling of the early QGP stage \cite{Bjorken:1982qr, Heinz:2013th}. At the LHC, the system starts at $T_0 \simeq 0.50$~GeV \cite{ALICE:2022} and cools down to $T \simeq 0.15$~GeV over a lifetime of $\tau_f \sim 10$~fm/$c$, whereas at RHIC the corresponding evolution proceeds from $T_0 \simeq 0.32$~GeV~\cite{PHENIX:2005} to freeze-out within $\tau_f \sim 7$~fm/$c$. The density profiles exhibit the expected 
Bjorken-like dilution \cite{Bjorken:1982qr}, with faster dilution at the LHC due to larger initial densities.  

Since both $T(\tau)$ and $n(\tau)$ decrease as the plasma expands, the threshold falls with time, defining a dynamical ``clock'' that governs whether the color field surpasses its breaking point. Hotter and denser initial conditions at the LHC make wave--breaking more likely at earlier times compared to RHIC.

In Fig.~\ref{fig:Ewb}, we compare the time evolution of the chromoelectric field $E(\tau)$ with the corresponding wave--breaking limit $E_{\text{WB}}(\tau)$. At early times, $E_{\text{WB}}$ dominates over the self--generated field. As the system expands, $E(\tau)$ increases relative to $E_{\text{WB}}$, leading to eventual crossing points that mark the onset of wave--breaking. For LHC conditions, the crossing occurs at $\tau_{\text{WB}}^{\text{LHC}} \simeq 2.3$~fm/$c$, while at RHIC it takes place slightly later at $\tau_{\text{WB}}^{\text{RHIC}} \simeq 2.8$~fm/$c$. The earlier onset at the LHC is a direct consequence of the higher initial temperature and density, which enhance collective plasma modes and reduce the 
effective wave--breaking threshold.  


The picture emerging from these results establish that plasma wave--breaking is expected well within the lifetime of the QGP in both RHIC and LHC collisions. The short timescales, compared to the overall hydrodynamic evolution, suggest that nonlinear plasma instabilities may play a significant role in the early pre-equilibrium stage of heavy-ion collisions. This 
provides a natural mechanism for energy dissipation and turbulent field generation, potentially impacting quarkonium suppression and jet quenching observables \cite{Mrowczynski:2017, Berges:2021}.

\section{summary and outlook}

In summary, we have analyzed the onset of wave--breaking in the expanding QGP background for both LHC Pb--Pb collisions at $\sqrt{s_{\rm NN}}=5.02$ TeV (ALICE conditions) and RHIC Au--Au collisions at $\sqrt{s_{\rm NN}}=200$ GeV. At the LHC, the higher initial temperature and density drive the nonlinear color field above the decreasing wave--breaking threshold within $\tau \lesssim $ 2.3 fm/$c$, while at RHIC the cooler, more dilute plasma delays the instability to $\tau \simeq 2.8$~fm/$c$. 
Parameter scans confirm that higher initial temperature, shorter thermalization time, and longer QGP lifetime accelerate the growth of unstable modes, while larger initial density tends to stabilize them. Consequently, the hotter and longer–lived plasma at the LHC exhibits earlier and stronger wave–-breaking than at RHIC. Despite these differences, the overall trend is robust: expanding QGP backgrounds inevitably lower the stability threshold and force coherent fields to break on timescales far shorter than kinetic equilibration would suggest. Wave--breaking thus provides a unified microscopic mechanism for the rapid loss of coherence and fast approach to hydrodynamic behavior observed across RHIC and LHC energies, linking initial chromodynamic field instabilities to the near--perfect fluidity of hot QCD matter.


The identification of wave--breaking as an early--time dynamical feature of the QGP has several implications for heavy--ion phenomenology. By driving nonlinear color--field decay within $\tau \lesssim 2.3$ fm/c at the LHC and slightly later at RHIC, these instabilities can accelerate plasma isotropization and enhance the transfer of field energy into hard and soft partonic modes. Such dynamics may affect jet energy loss, modify initial transport coefficients, and influence quarkonium suppression through fluctuating color backgrounds. 
In particular, the earlier onset of wave–breaking at the LHC suggests stronger field-induced modifications to early jet quenching and quarkonium regeneration processes than at RHIC~\cite{JetQuenchInitStage2019,QuarkoniumTransportReview2017}.


\section{DATA AVAILABILITY}
The data supporting this study’s findings are available
within the article.

\bibliographystyle{apsrev4-2}

\begin{thebibliography}{99}

\bibitem{Dawson:1959} J.~M.~Dawson,
Phys.\ Rev.\ {\bf 113}, 383 (1959).

\bibitem{Maity2013} C.~Maity \textit{et al.},
Phys. Rev. Lett. \textbf{110}, 215002 (2013).

\bibitem{Romatschke2003} P. Romatschke and M. Strickland, 
Phys. Rev. D \textbf{68}, 036004 (2003).

\bibitem{Gelis2010} F. Gelis, T. Lappi, and R. Venugopalan, 
Int. J. Mod. Phys. E \textbf{16}, 2595 (2010).

\bibitem{ALICE:2016ccg}
J.~Adam \textit{et al.} (ALICE Collaboration), 
Phys.\ Rev.\ Lett.\ \textbf{116}, 132302 (2016),

\bibitem{PHOBOS:2007dmu}
B.~Alver \textit{et al.} (PHOBOS Collaboration), 
Phys.\ Rev.\ Lett.\ \textbf{98}, 242302 (2007),

\bibitem{ALICE:2024}
S.~Acharya \textit{et al.} (ALICE Collaboration),
Phys.\ Rev.\ Lett.\ \textbf{132}, 042301 (2024).

\bibitem{ALICE:2016fzo}
J.~Adam \textit{et al.} (ALICE Collaboration), 
Phys.\ Lett.\ B \textbf{754}, 235--248 (2016),

\bibitem{PHENIX:2021nib}
U.~Acharya \textit{et al.} (PHENIX Collaboration), 
Phys.\ Rev.\ C \textbf{103}, 024904 (2021),

\bibitem{Bjorken:1982qr}
J.~D.~Bjorken, Phys.\ Rev.\ D {\bf 27}, 140 (1983).

\bibitem{Heinz:2013th}
U.~Heinz and R.~Snellings, Ann.\ Rev.\ Nucl.\ Part.\ Sci.\ {\bf 63}, 123 (2013).

\bibitem{Baier:2000mf}
R.~Baier, A.~H.~Mueller, D.~Schiff, and D.~T.~Son,
Phys.\ Lett.\ B {\bf 502}, 51 (2001).

\bibitem{Kurkela:2018qeb}
A.~Kurkela and A.~Mazeliauskas,
Phys.\ Rev.\ Lett.\ {\bf 122}, 142301 (2019).

\bibitem{Akhiezer:1975}
A.~I.~Akhiezer and R.~V.~Polovin, Pergamon Press (1975).

\bibitem{Blaizot:2001nr}
J.-P.~Blaizot and E.~Iancu,
Phys.\ Rept.\ {\bf 359}, 355 (2002).

\bibitem{Lebedev:1989ev}
V.~Lebedev and A.~Smilga,
Annals Phys.\ {\bf 202}, 229 (1990).


\bibitem{Mrowczynski:1993qm}
S.~Mr\'owczy\'nski,
Phys.\ Lett.\ B {\bf 314}, 118 (1993).

\bibitem{Romatschke:2003ms}
P.~Romatschke and M.~Strickland,
Phys.\ Rev.\ D {\bf 68}, 036004 (2003).






\bibitem{ALICE:2022wpn} S.~Acharya \textit{et al.}, ALICE Collaboration, JHEP \textbf{11}, 013 (2022).  
  
\bibitem{Gale:2013da} C.~Gale, S.~Jeon, and B.~Schenke, Int.\ J.\ Mod.\ Phys.\ A \textbf{28}, 1340011 (2013).  

\bibitem{PHENIX:2004vcz}S.~S.~Adler \textit{et al.}, PHENIX Collaboration, Phys.\ Rev.\ C \textbf{69}, 034909 (2004).  
\bibitem{STAR:2008med}B.~I.~Abelev \textit{et al.}, STAR Collaboration, Phys.\ Rev.\ C \textbf{79}, 034909 (2009).



\bibitem{ALICE:2022}
S.~Acharya \textit{et al.} (ALICE Collaboration),
Phys.\ Rev.\ Lett.\ \textbf{128}, 172301 (2022).

\bibitem{PHENIX:2005}
K.~Adcox \textit{et al.} (PHENIX Collaboration),
Nucl.\ Phys.\ A \textbf{757}, 184 (2005).

\bibitem{Mrowczynski:2017}
S.~Mrowczynski, B.~Schenke, and M.~Strickland,
Phys.\ Rept.\ \textbf{682}, 1 (2017).

\bibitem{Berges:2021}
J.~Berges, M.~P.~Heller, A.~Mazeliauskas, and R.~Venugopalan,
Rev.\ Mod.\ Phys.\ \textbf{93}, 035003 (2021).

\bibitem{JetQuenchInitStage2019}
C.~Andrés, N.~Armesto, H.~Niemi, R.~Paatelainen, and C.~A.~Salgado,
Phys.\ Rev.\ C \textbf{99}, 054910 (2019).

\bibitem{QuarkoniumTransportReview2017}
X.~Du and R.~Rapp,
Nucl.\ Phys.\ A \textbf{932}, 96 (2014).

\end{thebibliography}

\end{document}